\documentclass[conference]{IEEEtran}
\usepackage[left=0.75in,right=0.75in,top=1in,bottom=0.75in]{geometry}

\usepackage{graphicx}
\usepackage[cmex10]{amsmath}
\usepackage{amsfonts}
\usepackage[ruled]{algorithm2e}
\usepackage{todonotes}

\SetKwComment{Comment}{$\triangleright$\ }{}

\begin{document}
\title{Distributed Black-Box Optimization via Error Correcting Codes}

\author{
\IEEEauthorblockN{Burak Bartan and Mert Pilanci}
\IEEEauthorblockA{Department of Electrical Engineering, Stanford University\\
Email: {\{bbartan, pilanci\}@stanford.edu}
}}

\maketitle

\begin{abstract}
We introduce a novel distributed derivative-free optimization framework that is resilient to stragglers. The proposed method employs coded search directions at which the objective function is evaluated, and a decoding step to find the next iterate. Our framework can be seen as an extension of evolution strategies and structured exploration methods where structured search directions were utilized. As an application, we consider black-box adversarial attacks on deep convolutional neural networks. Our numerical experiments demonstrate a significant improvement in the computation times.
\end{abstract}

\section{Introduction}
Derivative-free optimization is an important computational task in many areas such as machine learning, statistics, design optimization and decision making \cite{conn2009introduction}. In many applications, it is common to encounter black-box optimization problems where the gradient of the objective function is not available. It is also possible that the gradient information is unreliable, or impractical to obtain. Derivative-free optimization methods employ only the function values to make progress toward an optimal solution. They can be used to directly tackle complex and diverse problems without an analytical form of the objective function. Additionally, since the function values can be evaluated independently in parallel, derivative-free methods can exploit parallelism to a great extent. 

In the recent years, cloud computing has offered inexpensive and scalable solutions to computational problems, and is widely adopted and used by the community. However, cloud-based serverless computing systems typically suffer from system noise, latency and variability in computation times \cite{dean2013tail,hoefler2010characterizing}. This results in a subset of slower worker nodes called stragglers.  In this work, we propose a distributed derivative-free optimization method that is resilient to stragglers. Our method forms gradient estimates by evaluating the function at several points by employing coding theoretic ideas and tools. We first encode ordinary basis perturbations using Hadamard transform into perturbation directions along the rows of the Hadamard matrix $H$. The objective function is evaluated at the encoded points in parallel, and then a gradient direction is determined from the available workers via decoding. We establish a connection between derivative free optimization via the well-known finite differences method, recently proposed structured evolution strategies \cite{choromanski2018structured}, and polar codes \cite{polar2009arikan}. The structured evolution strategies approach corresponds to averaging the outputs of function evaluations with variables perturbed along the rows of $HD$ where $H$ is the Hadamard matrix and $D$ is a diagonal matrix with random diagonal entries. When the decoding step is omitted, encoding the unit perturbations leads to the structured evolution strategies which is simple to implement. If we decode the outputs instead of averaging them using the modified successive cancellation decoder of polar codes, however, we obtain partial derivatives as in the finite differences method. Because the function evaluations are scalar valued and the decoder can run in near linear time, the time it takes to decode these scalar outputs is negligible. For a large class of problems, including the decoding step yields faster convergence in optimization. The proposed method also enables a way to switch between finite differences and structured evolution strategies and provide a more favorable trade-off.

%


\subsection{Related Work}
This work draws ideas from coded computation literature where the goal is to speed up distributed matrix multiplication. The authors in \cite{Lee2018} introduced the concept of using Maximum Distance Separable (MDS) codes to compute matrix-vector multiplications faster in distributed computing platforms. The way coding is integrated into the computation of a matrix vector product $Ax$, where $A\in \mathbb{R}^{n\times d}$ and $x\in\mathbb{R}^d$ is as follows: First $A$ is partitioned into submatrices along the rows and then these submatrices are encoded using MDS codes. The workers multiply their assigned submatrix with $x$. Because the encoded $A$ is redundant, it is possible to decode $Ax$ via a standard decoder without accessing the outputs of all workers. Works that introduce variants and extensions of this idea include \cite{severinson2018lt}, \cite{mallick2018rateless} and \cite{polar_computation}. There are many other recent studies on coded computation that explore the partitioning of both matrices being multiplied, e.g., \cite{yu2017polycode}, \cite{yu2018straggler}. This work is more related to the case where only one of the matrices is partitioned.

Another line of study that this work is connected to is the study of black-box optimization methods for reinforcement learning. The authors in \cite{salimans2017es} explore the use of the evolution strategies method in reinforcement learning and show that distributed training with evolution strategies can be very fast because of its scalability. The work in \cite{choromanski2018structured} shows that using orthogonal exploration directions leads to lower errors and present the \textit{structured} evolution strategies method which is based on a special way of generating random orthogonal exploration directions, as we discuss later in detail.

\newgeometry{top=0.75in,bottom=0.75in,right=0.75in,left=0.75in}
\subsection{Main Contributions}
We introduced a novel framework based on coded gradient estimates that leads to straggler-resilient distributed optimization of black-box functions. Our method introduces a novel connection between the finite differences method and the structured evolution strategies method by using polar codes \cite{polar2009arikan}. Once function evaluations at perturbed parameters are returned, it is possible to use them as they are without decoding, which is equivalent to the structured evolution strategies method or decode them to obtain the gradient estimate and use it in a gradient-based optimization algorithm. Our method offers this flexibility at the expense of doing some additional inexpensive computing for decoding because of the efficient polar code decoder.

\section{Preliminaries}
\subsection{Notation}
We are interested in solving the black-box optimization problem given by
\begin{equation*}
\begin{aligned}
& \underset{\theta}{\text{minimize}}
& & f(\theta)
\end{aligned}
\end{equation*}
where $f : \mathbb{R}^d \to \mathbb{R}$. Throughout the text, unless otherwise stated, we assume that we don't have access to the analytical form of $f(\theta)$ or its gradient, and we assume that we are only able to make queries for function evaluations. We use $\nabla_v$ to denote directional derivative along the direction $v$. We use $e_i$ to denote the $i$'th unit column vector with the appropriate dimension. Throughout the text, $H$ refers to the Hadamard matrix (its dimension can be determined from the context), and $\mathbf{H}$ refers to the Hessian of a function.

\subsection{Finite Differences}
For black-box optimization problems, approximate gradients can be obtained by using the finite differences estimator. The derivative of a function $f(x)$ with respect to the variable $x_i$ can be approximated by using
\begin{align}\label{finite_diff}
    \frac{\partial f(x)}{\partial x_i} \approx \frac{f(x+\delta e_i) - f(x-\delta e_i)}{2\delta},
\end{align}
where $e_i$ is the $i$'th unit vector and $\delta \in \mathbb{R}$ is a small scalar that determines the perturbation amount. One can obtain an approximate gradient using \eqref{finite_diff} and feed these estimate gradients to gradient-based optimization methods such as gradient descent.

The finite differences estimator can easily exploit parallelism since the partial approximate derivatives can be computed in different workers in a distributed setting. This method can be modified to be straggler-resilient by using only the available derivative estimates and ignoring the outputs of the slower workers. However, this typically leads to slower convergence. We note that the proposed method results in a mechanism that always returns all entries of the gradient estimate and is straggler-resilient at the same time. Another alternative is to replicate finite difference calculations, which is not optimal from a coding theory perspective.

\subsection{Evolution Strategies}
Let us consider the following evolution strategies (ES) gradient estimator (\cite{salimans2017es}, \cite{choromanski2018structured})
\begin{align}\label{ES_estimator}
    \nabla f(\theta) \approx \frac{1}{2N\delta} \sum_{i=1}^{N}(f(\theta+\delta \epsilon_i)\epsilon_i - f(\theta-\delta \epsilon_i)\epsilon_i)
\end{align}
where $\delta$ is the scaling coefficient for the random perturbation directions $\epsilon_i$ and $N$ is the number of perturbations. This estimator is referred to as antithetic evolution strategies gradient estimator. There are other gradient estimators slightly different from the one in \eqref{ES_estimator}. In this work, we always consider the antithetic version given in \eqref{ES_estimator} because it leads to better empirical results in our numerical simulations. The random perturbation directions $\epsilon_i$ may be sampled from a standard multivariate Gaussian distribution $\mathcal{N}(0,I)$. Alternatively, $\epsilon_i$'s may be generated as the rows of $HD$ where the $D$ is a diagonal matrix with entries distributed according to Rademacher distribution. The authors in \cite{choromanski2018structured} propose generating perturbation directions by repetitive multiplications $HD_1...HD_k$ where $D_i$'s are independent. 

It is shown in \cite{choromanski2018structured} that if exploration (or perturbation) directions $\epsilon_i$ are orthogonal, the gradient estimators lead to lower error. We note that the rows of $HD$ are orthogonal with the appropriate scaling factor. We omit this scaling factor by absorbing it in the $\delta$ term which scales perturbation directions $\epsilon_i$. This work considers the case where random perturbation directions are generated according to $HD$ and we will refer to it as \textit{structured evolution strategies} as they do in \cite{choromanski2018structured}.

Evolution strategies can also exploit parallelism as well since workers need to communicate only scalars which are the function evaluations and the random seeds used when generating the random perturbation directions.

\subsection{Coded Computation}
Coded computation has recently been introduced in \cite{Lee2018} for speeding up distributed matrix multiplication. The idea is to add redundancy to computations by using linear codes in order to make it possible to recover the output of the multiplication without having all worker outputs. The fact that matrix multiplication is a linear operation plays a key role in coded computation. The scheme in \cite{Lee2018} is based on MDS codes. More recently, the authors in \cite{polar_computation} proposed a coded computation mechanism for serverless computing platforms using polar codes instead of MDS codes. Polar codes have low encoding and decoding complexity $O(N\log N)$.

This work draws ideas from the coded computation literature. In coded computation, for computing $Ax$, one encodes submatrices $A_i$ and decodes worker outputs $A_ix$. In coded optimization, the perturbation directions are encoded and function evaluations are decoded.

Many of the coded computation schemes have the restriction that they are designed to work with finite field data and not applicable to full-precision inputs. The decoding algorithm introduced in \cite{polar_computation} for polar codes can however work with full-precision data. Since we do not restrict the output of the black-box function $f$ to take values in a finite field, we need a decoder that can operate on full-precision real values. Quantization is an option, though it would introduce noise, which can be avoided by using a decoding algorithm that can work with full-precision data. We note that it is possible to use the decoder in \cite{polar_computation} with a slight modification to account for the difference in the encoding kernels ($\bigl[\begin{smallmatrix} 1 & 1 \\ 0 & 1 \end{smallmatrix}\bigr]$ for polar codes, and $\bigl[\begin{smallmatrix} 1 & ~\,\,1 \\ 1 & -1 \end{smallmatrix}\bigr]$ for the Hadamard transform). We discuss the decoding in detail in the next section.

\section{Coded Black-Box Optimization}
\begin{figure}[htbp]
  \centering
  \includegraphics[width=0.45\textwidth]{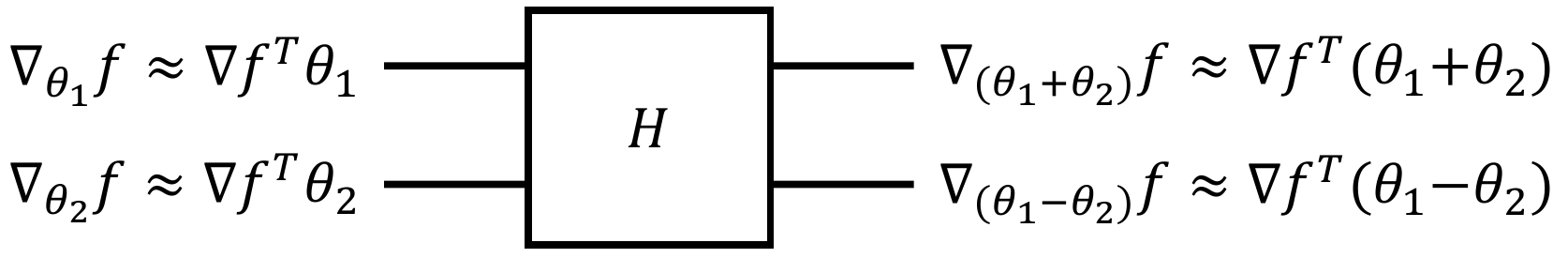}
  \caption{2-by-2 construction based on Hadamard transformation.}
  \label{example_construction}
\end{figure}

In this section, we present the proposed method for speeding up distributed black-box optimization. We start by introducing some more notation and definitions that we will be using. The derivative of a differentiable function $f$ at a point $\theta$ along the unit vector direction $e_i$ is the $i$'th component of the gradient $\nabla f$. More precisely, we have
\begin{align}
    \frac{\partial f}{\partial \theta_i} = e_i^T \nabla f.
\end{align}
The directional derivative along $v$ is defined as follows
\begin{align}\label{direc_deriv_2}
    \nabla_{v}f = \lim_{\delta \rightarrow 0} \frac{f(\theta + \delta v) - f(\theta) }{\delta}\,.
\end{align}
If the function is differentiable at a point $\theta$, then the directional derivative exists along any direction $v$ and is a linear map \cite{Boyd02}. In this case we have
\begin{align}\label{direc_deriv_1}
    \nabla_{v}f = v^T \nabla f.
\end{align}
When we do not have access to exact gradients, we can employ a \emph{numerical directional derivative} by choosing a small $\delta$ in
\begin{align}\label{direc_deriv_approx}
    \nabla_{v}f \approx \frac{f(\theta + \delta v) - f(\theta) }{\delta}.
\end{align}
Note that the approximation in \eqref{direc_deriv_approx} is not symmetric, that is, we only perturb the parameters along $+v$. We instead use the symmetric version of \eqref{direc_deriv_approx} for approximating derivatives in which the parameters are perturbed along both the directions $-v$ and $+v$:
\begin{align}\label{direc_deriv_symmetric}
    \nabla_{v}f \approx \frac{f(\theta + \delta v) - f(\theta - \delta v) }{2\delta}.
\end{align}
If we consider the Taylor series expansion for $f(\theta + \delta v)$, we obtain
\begin{align}
    f(\theta + \delta v) &= f(\theta) + \delta \nabla f^T v + \frac{\delta^2}{2} v^T \mathbf{H}v + \mathcal{O}(\delta^3)
\end{align}
and for $f(\theta - \delta v)$, we have
\begin{align}
    f(\theta - \delta v) &= f(\theta) - \delta \nabla f^T v + \frac{\delta^2}{2} v^T \mathbf{H}v + \mathcal{O}(\delta^3)
\end{align}
where $\mathcal{O}(\delta^3)$ is a third order error term and $\mathbf{H}$ is the Hessian matrix for $f$. Substituting these expansions in \eqref{direc_deriv_symmetric}, we obtain
\begin{align}\label{direc_deriv_final}
    \nabla_{v}f &\approx \frac{ 2\delta \nabla f^T v + \mathcal{O}(\delta^3) }{2\delta} \nonumber \\
    &= \nabla f^T v + \mathcal{O}(\delta^2).
\end{align}
This shows that by choosing $\delta$ small, we can make the numerical directional derivative approximately linear in $\nabla f$. Our proposed method makes use of this assumption that gradient estimates are linear in their directions $v$ to ensure that coding can be applied to directional derivative estimates. To make this more concrete, let us consider the construction given in Figure \ref{example_construction}. The block $H$ simply corresponds to the linear transformation $H=\bigl[\begin{smallmatrix} 1 & 1 \\ 1 & -1 \end{smallmatrix}\bigr]$.

In Figure \ref{example_construction}, if we know the estimates for $\nabla_{(\theta_1+\theta_2)}f$ and $\nabla_{(\theta_1-\theta_2)}f$, we can compute the estimates for $\nabla_{\theta_1}f$ and $\nabla_{\theta_2}f$ because the directional derivative estimates are approximately linear in their corresponding directions. Furthermore, if we know the estimate for $\nabla_{\theta_1}f$, then it is sufficient to know only one of the estimates for $\nabla_{(\theta_1+\theta_2)}f$ or $\nabla_{(\theta_1-\theta_2)}f$ in order to be able to compute the estimate for $\nabla_{\theta_2}f$. This would happen if, for example, $\theta_1$ is the zero vector (i.e. frozen direction) because the estimate for $\nabla_{\theta_1}f$ would be zero and either of the directional derivative estimates from the right-hand side would be enough for us to obtain the estimate of $\nabla_{\theta_2}f$.

We refer to the directions $\theta_1$, $\theta_2$, $(\theta_1+\theta_2)$, $(\theta_1-\theta_2)$ as perturbation directions.

We now summarize the proposed method and detailed description of each item will follow:
\begin{itemize}
    \item Encode all the unit vectors in $\mathbb{R}^d$ to obtain the encoded perturbation directions.
    \item Assign each perturbation direction to a worker and have workers compute their directional derivative estimates using \eqref{direc_deriv_symmetric}.
    \item Master starts collecting worker outputs. 
    \item When a decodable set of worker outputs (i.e. directional derivative estimates) is available, the master node decodes these outputs to obtain an estimate for the gradient.
    \item The master node then makes a gradient update to find the next iterate for the parameter $\theta$.
    \item Repeat until convergence or for a desired number of iterations.
\end{itemize}
The above procedure assumes that we want to estimate all entries of the gradient, but it is possible estimate only a portion of gradient entries by encoding only the unit vectors corresponding to the desired entries. 

We note that we can always check whether decoding helps in obtaining a better objective function compared to the the structured evolution strategies and make the update accordingly. Because the decoding step is quite fast thanks to the polar decoder, having decoded the outputs but ended up not using the decoded estimate is not a computational burden.  

\subsection{Encoding}
\begin{figure}[htbp]
  \centering
  \includegraphics[width=0.23\textwidth]{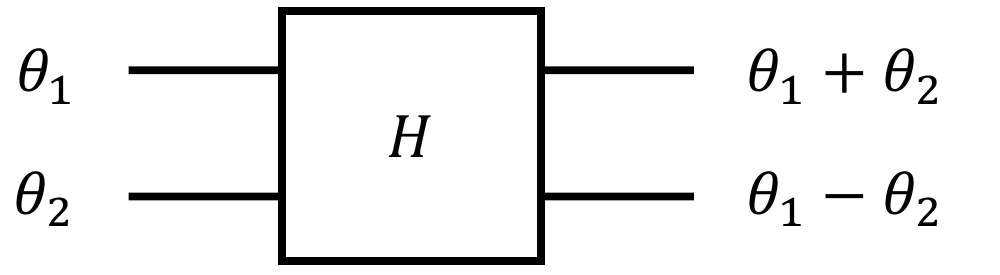}
  \caption{2-by-2 Hadamard transformation of the perturbation directions.}
  \label{two_by_two_perturb}
\end{figure}

Encoding is done based on the Hadamard transformation whose kernel is shown in Figure \ref{two_by_two_perturb}. Encoding is done the same way as polar encoding is done. Hence, the computational complexity of encoding is only $N\log N$ operations.

The way polar coding for the erasure channel works is that one computes the erasure probabilities of the transformed channels and freezes the channels with the highest erasure probabilities and the remaining channels are used for information. We refer the reader to \cite{polar2009arikan} and \cite{polar_computation} for more details on polar codes. In channel coding, freezing channels corresponds to sending known bits, e.g., zeros, at the receiver. In this work, freezing channels corresponds to setting it to zero coordinates such that the directional derivative along these vectors is zero. This makes it possible, when decoding, to take the value of the derivative estimates for frozen channels to be zero. For the information channels, we simply send in unit vectors.

\begin{figure}[htbp]
  \centering
  \includegraphics[width=0.48\textwidth]{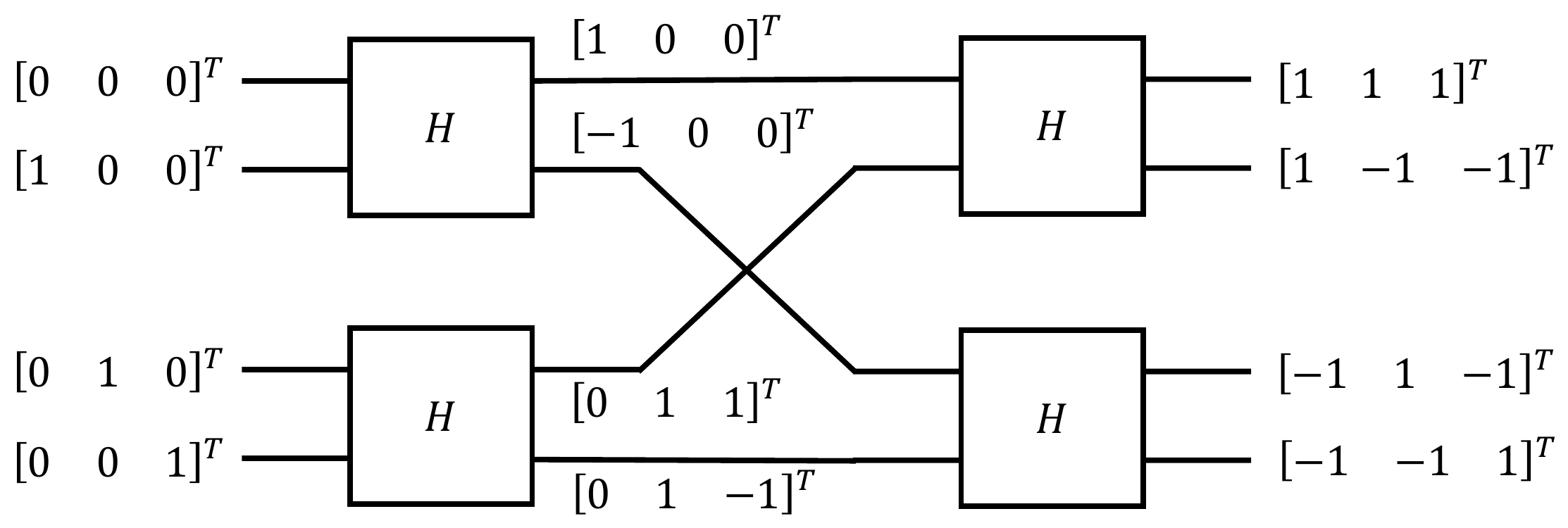}
  \caption{Example encoding.}
  \label{example_encoding}
\end{figure}

Encoding for a function of $3$ variables using $N=4$ workers is shown in Figure \ref{example_encoding}. The resulting $4$ output vectors are the perturbation directions.
In this construction, the rate is $\frac{3}{4}$ since $3$ out of $4$ channels are used for sending in unit vectors. When the erasure probabilities of the transformed channels are computed (see \cite{polar2009arikan}), one will see that the worst one corresponds to the first index. Hence, the first channel is the frozen channel and the remaining $3$ channels are the information channels. The frozen channel is set to the zero vector and the information channels are set to unit vectors. Sending in the zero vector for the frozen channel makes sense because perturbing by the zero vector is the same as not perturbing the variables and hence we get $f(\theta + 0) - f(\theta - 0) = 0$. This is important because during decoding, we will not have to do any computations to evaluate the value of the frozen inputs as we know they are zero. 

If we wish to use $N=8$ workers instead of $4$, we would set the worst $5$ channels to zero vectors and the remaining best three channels would be set to the unit vectors $e_1, e_2, e_3$. The rate in this case would be $\frac{3}{8}$, and this construction would be more straggler-resilient since we can recover the gradient estimate in the presence of even more stragglers compared to the $N=4$ case.

\textbf{Embedding interpretation:} It is also possible to perform the encoding step slightly differently for a different view on freezing channels. Instead of setting frozen channels to zero vectors, one can increase the dimension of the inputs from $d$ to $N$ and set the frozen channels to unit vectors $e_j$ where $j \in \{d+1,d+2,...,N\}$. Because the function $f(\theta)$ accepts $d$-dimensional inputs, we could embed $f(\theta)$ into a higher dimension, that is, we could define $\tilde{f}(\tilde{\theta})$ where $\tilde{\theta} \in \mathbb{R}^N$ and $\tilde{f}(\tilde{\theta}) = f(\theta)$ if $\tilde{\theta}_i=\theta_i$ for $i \in \{1,2,...,d\}$. The advantage of this approach is that the output of the encoding step will be equal to the $H$ matrix with permuted rows.

\subsection{Decoding}
\begin{figure}[htbp]
  \centering
  \includegraphics[width=0.44\textwidth]{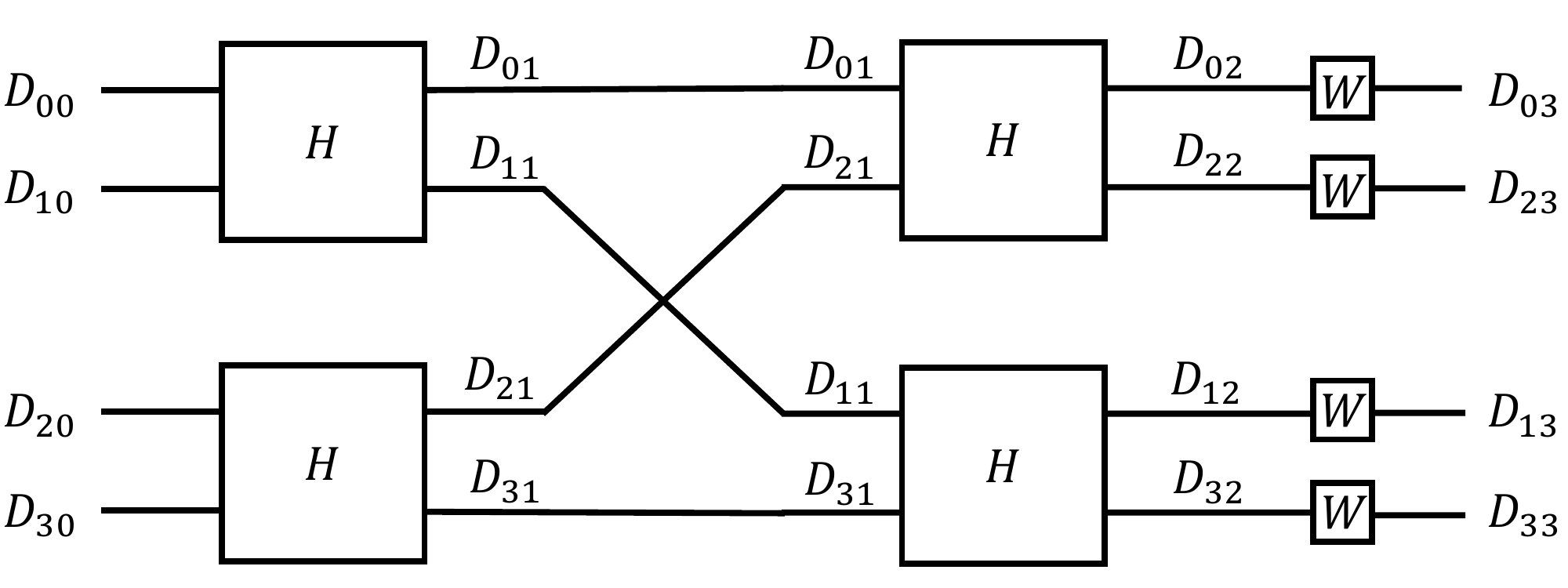}
  \caption{$N=4$ construction.}
  \label{example_encoding_for_notation}
\end{figure}

The sequential polar code decoder given in \cite{polar_computation} works for linear operations such as matrix-vector multiplication with full-precision data (i.e. does not require finite field data). Because we assume that the gradient estimates $\nabla_{h_i}f$ are approximately linear, we use the same decoder in \cite{polar_computation} by only modifying it slightly so that it decodes according to the Hadamard kernel and not the polar code kernel. The decoder is given in Alg. \ref{decoding_alg} along with the subroutine \textit{decodeRecursive(i,j)} in Alg. \ref{decode_rec}. The decoder uses the notation $D_{ij}$ to denote the value corresponding to the node $i$ in the $j$'th level in the coding circuit. Figure \ref{example_encoding_for_notation} shows this notation in the coding circuit for $N=4$. The blocks $W$ in Figure \ref{example_encoding_for_notation} represent workers. The notation $I_{D_{ij}}$ in Alg. \ref{decoding_alg} indicates whether the data for the node $i$ in the $j$'th level is available or erased (i.e., not computed yet). checkDecodability is another subroutine that checks whether it is possible to decode the available outputs given the indicators of output availabilities. It checks decodability of outputs by running the decoder on the indicators and if decoding fails, it returns false, otherwise, it returns true.

\DontPrintSemicolon
\begin{algorithm}
 \KwIn{Indices of the frozen channels}
 \KwResult{estimate of $\nabla f$ \Comment*[r]{Part I}}
 
 Initialize $I_{D_{:,0}} = [I_{D_{0,0}}, I_{D_{1,0}}, \dots, I_{D_{N-1,0}}] = [0,\dots,0]$ \;
 \While{$I_{D_{:,0}}$ not decodable}{
 update $I_{D_{:,0}}$\;
 checkDecodability($I_{D_{:,0}}$)
 }
 Initialize an empty list $y$  \Comment*[r]{Part II}
 \For{$i \gets 0$ \textbf{to} $N-1$} {
   $D_{i,0}$ = decodeRecursive($i$, $0$)\;
   \If{node $i$ is a data node} {
     $y = [y; D_{i,0}]$\;
   }
   \Comment*[r]{forward prop}
   \vspace{-0.3cm}
   \If{$i \bmod 2 = 1$ } {
     \For{$j \gets 0$ \textbf{to} $\log_2{N}$} {
       \For{$l \gets 0$ \textbf{to} $i$}{
         compute $D_{lj}$ if unknown 
       }
     }
   }
  }
  return $y$
 \caption{Decoding algorithm.}
 \label{decoding_alg}
\end{algorithm}

\begin{algorithm}
 \KwIn{Node $i \in [0,N-1]$, level $j \in [0,\log_2{N}]$}
 \KwResult{$I_{D_{ij}}$ and modifies $D$}
 
  \lIf{$j = \log_2{N}$} { 
     return $I_{D_{i,\log_2{N}}}$ \Comment*[r]{base case 1}
   }
   \vspace{-0.4cm}
  \lIf{$I_{D_{ij}} = 1$} {
     return $1$ \Comment*[r]{base case 2}
   }
   \vspace{-0.4cm}
  $I_{D_{i,(j+1)}} = $ decodeRecursive($i, j+1$)\;
  $I_{D_{pair(i),(j+1)}} = $ decodeRecursive($pair(i), j+1$)\;
 
  \uIf{$i$ is upper node}{
    \If{$I_{D_{i,(j+1)}} \text{ AND } I_{D_{pair(i),(j+1)}} = 1$}{
      compute $D_{ij}$\;
      return 1
    }
  }
  \Else{
    \If{$I_{D_{i,(j+1)}} \text{ OR } I_{D_{pair(i),(j+1)}}$ = 1}{
      compute $D_{ij}$\;
      return 1
    }
  }
 return 0
 \caption{decodeRecursive($i$, $j$)}
 \label{decode_rec}
\end{algorithm}
\subsection{Multiplication by $D$} 
Structured evolution strategies method is usually based on exploring the parameter space along the rows of $HD$ instead of $H$ alone. $D$ is a diagonal matrix with random diagonal entries distributed according to Rademacher distribution ($\pm 1$ with probability $0.5$). So far, we have only considered the $H$ matrix for perturbation directions. It is possible to incorporate the diagonal matrix $D$ into our method as well.

Multiplying $H$ by $D$ from the right corresponds to multiplying all the entries of the $i$'th column of $H$ by $D_{ii}$ for all $i$. In this case, instead of computing an estimate for the directional derivative along a direction of $v$, we will approximate the directional derivative with respect to the direction $Dv$. The derivations we have done for $v$ also apply for these randomized directions $Dv$. Decoding will work the same as before, except, after the decoding algorithm is run on the outputs, we need to divide the $i$'th entry of the decoded gradient estimate by $D_{ii}$ to obtain the entries of the gradient estimate.

\section{Speeding Up Distributed Computation of Adversarial Attacks}
Deep neural networks have found success in many different tasks. Their robustness is a critical performance criterion for many applications. In the literature, there are many works that propose methods to generate adversarial inputs to attack neural networks to investigate their robustness. Attack settings may differ in terms of levels of knowledge on the neural network model. Some works assume knowledge on the architecture of the neural network and this setting is referred to as open-box setting. On the other hand, if we only assume we have access to the outputs of the neural network, then this is called the black-box setting. In this section, we are interested in applying our proposed method to the black-box setting.

We apply our proposed method to the setting of finding adversarial examples based on black-box optimization. We adopt the settings in \cite{chen2017zoo_attacks} where they develop black-box attacks to neural network based classifiers and the goal is to find a misclassified image that looks like a given image $\theta_0$. They pose the following optimization problem for finding adversarial examples: minimize $f(\theta)$, where
\begin{align}\label{f_targeted}
    f(\theta) = ||\theta-\theta_0||_2^2 + c \max\{\max_{i \neq t}[F(\theta)]_i - \log[F(\theta)]_t, -\kappa \}.
\end{align}
$F(\theta)$ represents the class probabilities of the neural network when we feed in $\theta$ as the input. The subscript $i$ in $[F(\theta)]_i$ means the class probability for the $i$'th class. $c$ and $\kappa$ are non-negative parameters for tuning the attack. Note that the objective in \eqref{f_targeted} is for a targeted attack, meaning we wish to find an image that is classified as class $t$. If one is interested in finding an example misclassified but not necessarily to a given class, then this is called an untargeted attack. The optimization problem to be solved for untargeted attacks is given in \cite{chen2017zoo_attacks} as
\begin{align}\label{f_untargeted}
    f(\theta) &= \nonumber \\
    & ||\theta-\theta_0||_2^2 + c \max\{\log [F(\theta)]_{t_0}-\max_{i\neq t_0} \log [F
    (\theta)]_i, -\kappa \}
\end{align}
where $t_0$ is the correct class for the image that we wish to misclassify. Note that since we assume we have no information about the neural network $F(\theta)$, we cannot compute the gradient of the functions in \eqref{f_targeted} and \eqref{f_untargeted}. 

The proposed approach in \cite{chen2017zoo_attacks} involves selecting a given number of parameters in every iteration and estimating the derivatives only for these parameters. Hence, in every iteration, only a number of selected parameters are updated. The updates are done using the ADAM algorithm \cite{adam2014}. We present results regarding the use of our method in black-box adversarial attacks problem in numerical results section.
\section{Numerical Results}
We now present results showing the performance of our proposed framework.


\subsection{An $l_1$-norm Based Objective as the Black-Box Function}
In this subsection, we present experiment results for optimizing a known objective function $f(\theta) = ||A\theta - b||_1$ as a test case. Even though we experiment with a known function, we still use the black-box optimization algorithms assuming we do not have the exact gradients. 


\begin{figure}[htbp]
  \centering
  \includegraphics[width=0.3\textwidth]{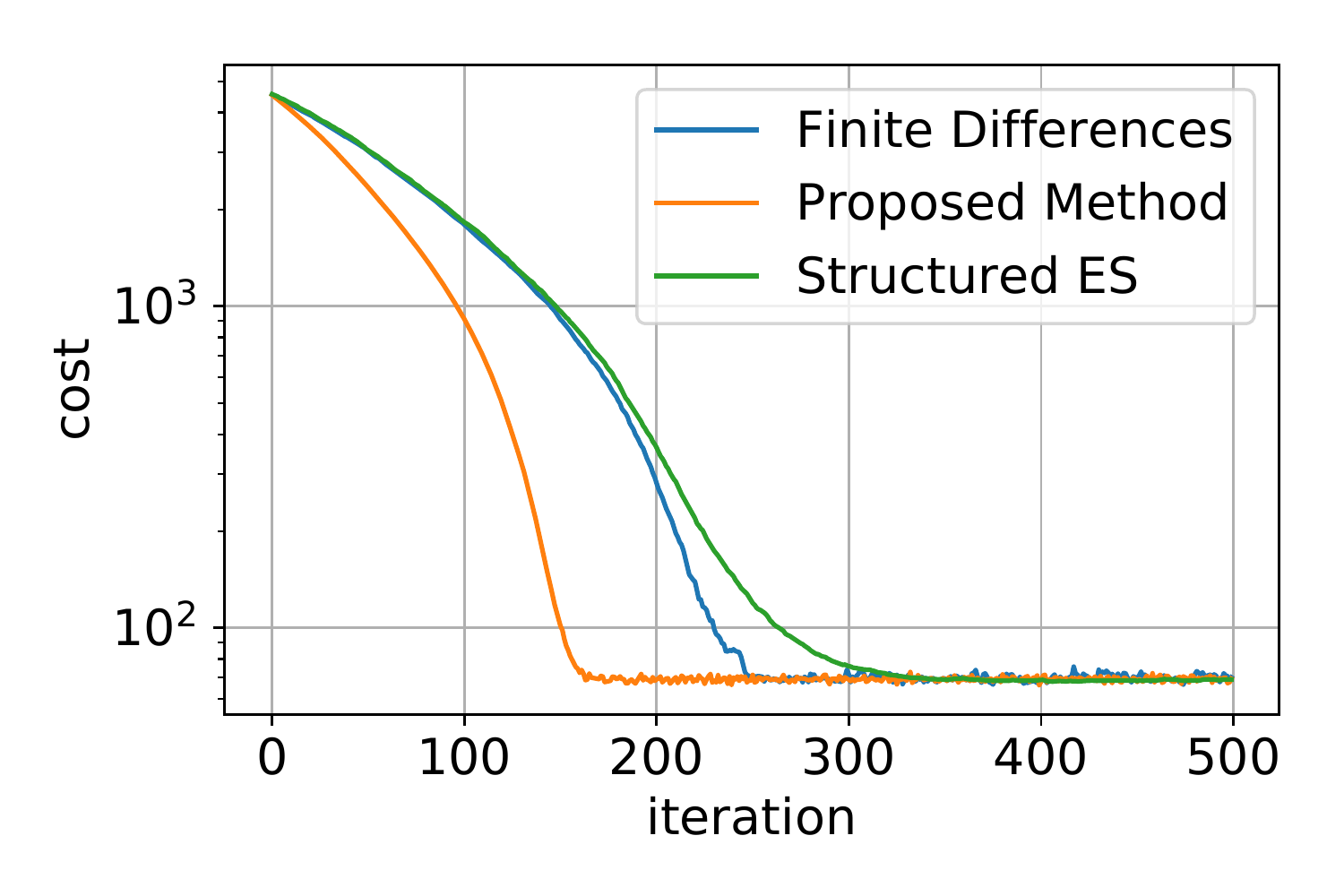}
  \caption{We optimize the function $f(\theta)=||A\theta-b||_1$ where $A \in \mathbb{R}^{200\times 32}$ and $y \in \mathbb{R}^{200}$ using black-box optimization methods.}
  \label{l1_norm_example_comparison}
\end{figure}

Figure \ref{l1_norm_example_comparison} shows the cost as a function of iterations when we use gradient descent algorithm with gradient estimates obtained by the finite differences method, the proposed method, and the structured evolution strategies method. For fairness in comparing these methods, we used straggler-resilient versions of the finite differences method and the structured evolution strategies method in obtaining these results. To clarify, for the finite differences method, we wait for the first arriving $16$ worker outputs out of the total $32$ outputs to make a gradient update. For the structured evolution strategies method, the perturbation directions come from the rows of $HD$ matrix which is a $32\times 32$ matrix, so we wait for the first arriving $16$ outputs out of the $32$. Finally, the way we implement the proposed method is that we use a rate of $\frac{1}{2}$ with a total of $64$ workers and wait for the first decodable set of outputs out of a total of $64$ outputs. Figure \ref{l1_norm_example_comparison} illustrates that having all the entries of the gradient estimate through decoding leads to faster convergence than having only a half of the entries of the gradient estimate. It also shows that the proposed method results in faster convergence compared to the structured evolution strategies method. 


\subsection{Black-Box Adversarial Attacks}
We now present some experiment results for our proposed method applied to finding adversarial examples. We have tested our proposed method for solving the optimization problem for targeted black-box attacks (objective of which is given in \eqref{f_targeted}) to a deep learning model. The model we considered is the VGG-16 architecture \cite{vgg16} trained on CIFAR-10 dataset \cite{cifar10_dataset} for image classification. VGG-16 architecture is a 16-layer convolutional neural network. CIFAR-10 is an image dataset with a total of 60000 images of size $32\times32\times3$ (50000 training and 10000 test images) with 6000 images per class for a total of 10 classes.

\begin{figure}[htb]
\begin{minipage}[b]{0.49\linewidth}
  \centering
  \centerline{\includegraphics[width=4cm]{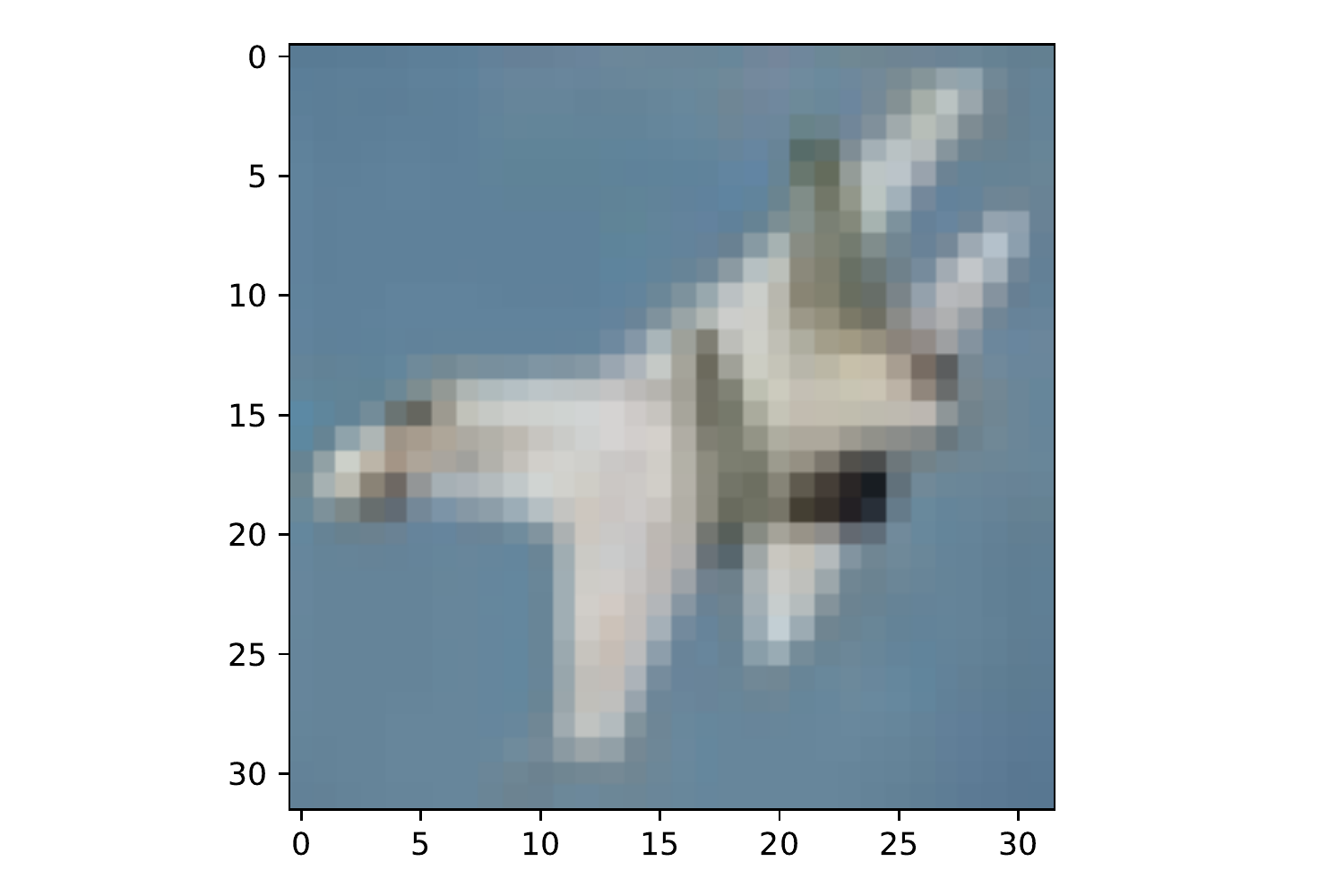}}
  \centerline{(a) Airplane classified correctly}\medskip
\end{minipage}
\hfill
\begin{minipage}[b]{0.49\linewidth}
  \centering
  \centerline{\includegraphics[width=4cm]{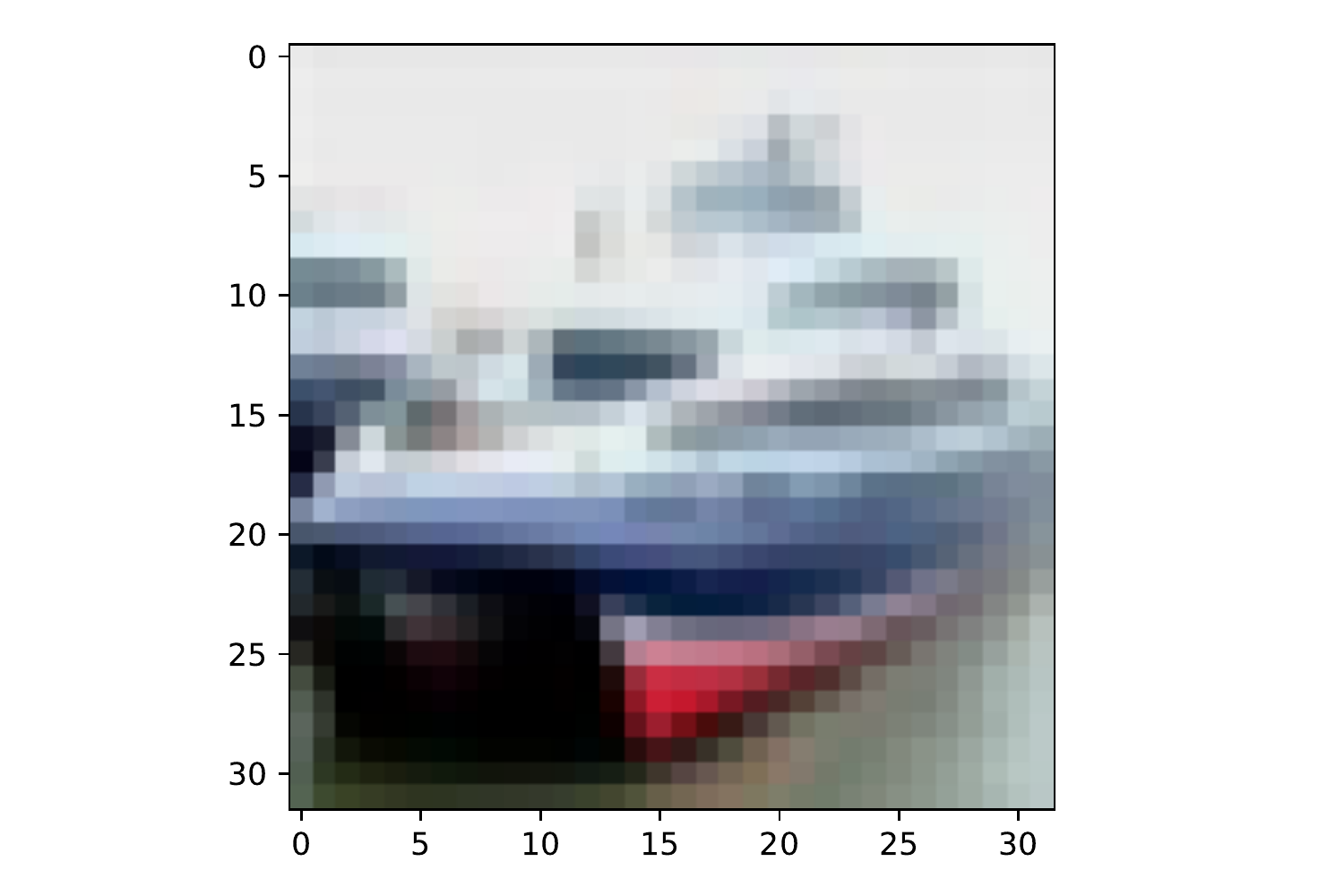}}
  \centerline{(b) Ship classified correctly}\medskip
\end{minipage}
\caption{Original images $\theta_0$.}
\label{original_images}
\end{figure}

\begin{figure}[htb]
\begin{minipage}[b]{0.49\linewidth}
  \centering
  \centerline{\includegraphics[width=4cm]{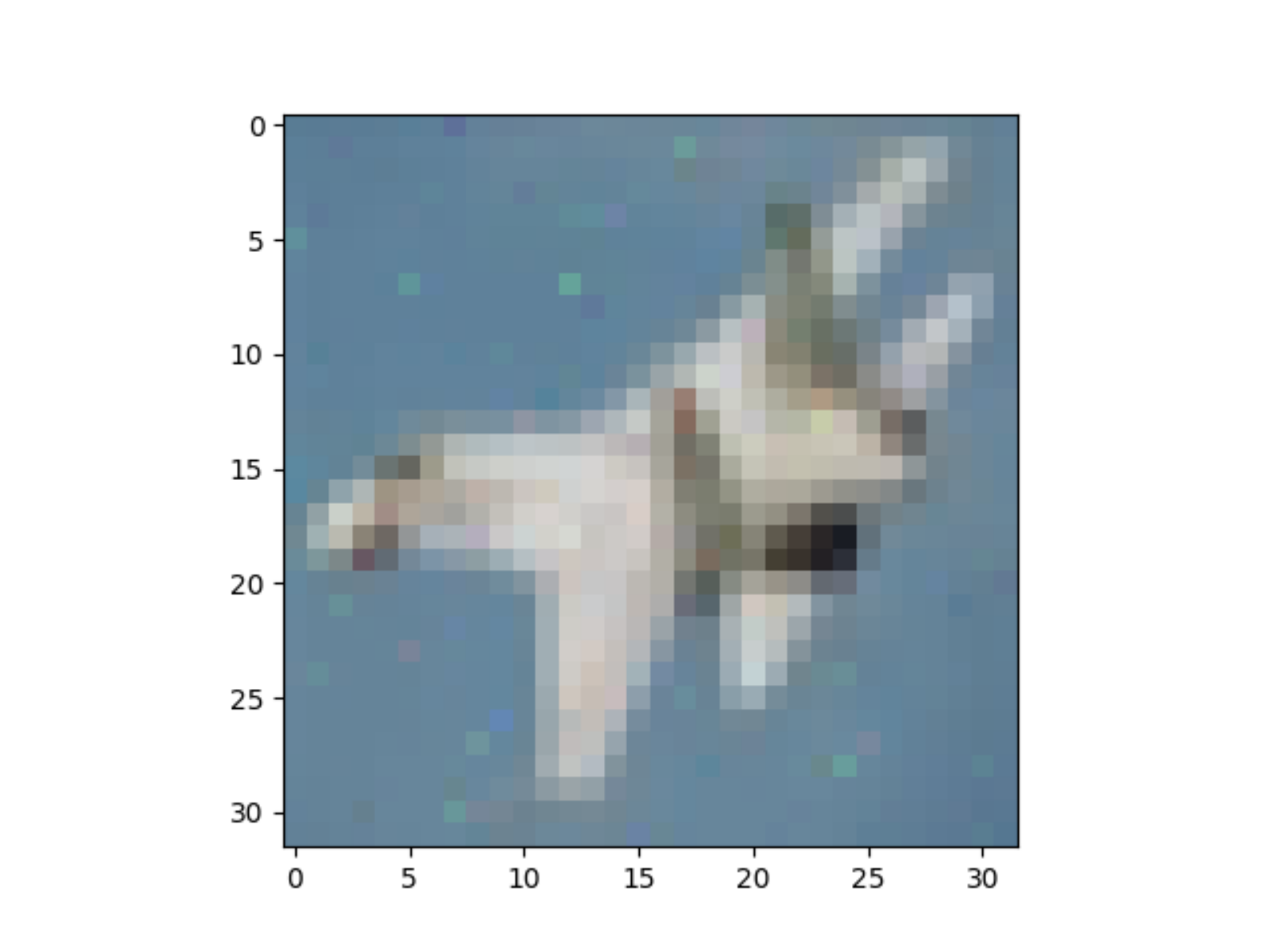}}
  \centerline{(a) Airplane classified as dog} \medskip
\end{minipage}
\hfill
\begin{minipage}[b]{0.49\linewidth}
  \centering
  \centerline{\includegraphics[width=4cm]{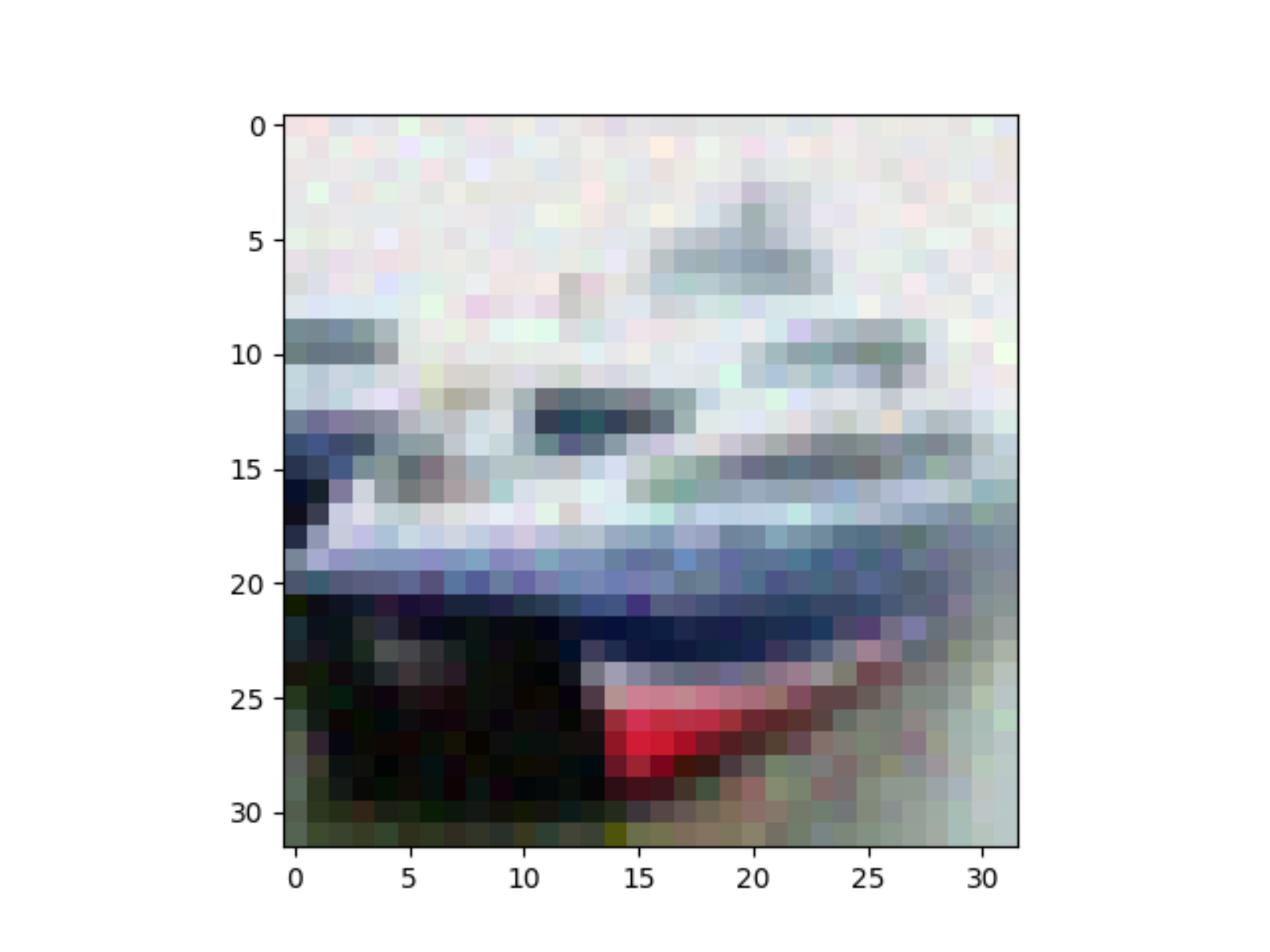}}
  \centerline{(b) Ship classified as dog}\medskip
\end{minipage}
\caption{Adversarial images generated using the proposed method.}
\label{adversarial_images}
\end{figure}

Before we delve into the details of the experiments, we show two original input images from CIFAR-10 and the generated adversarial images that look almost the same as the original images but are misclassified. Figure \ref{original_images} shows the original images $\theta_0$ which are correctly classified by the neural network $F(\theta)$. Figure \ref{adversarial_images} shows the generated adversarial images obtained by solving the black-box optimization problem for targeted attacks where the target class is the dog class. The adversarial images in Figure \ref{adversarial_images} are both misclassified as dog. These adversarial images were obtained by using our proposed method.

\begin{figure}[htb]
\begin{minipage}[b]{0.47\linewidth}
  \centering
  \centerline{\includegraphics[width=4.5cm]{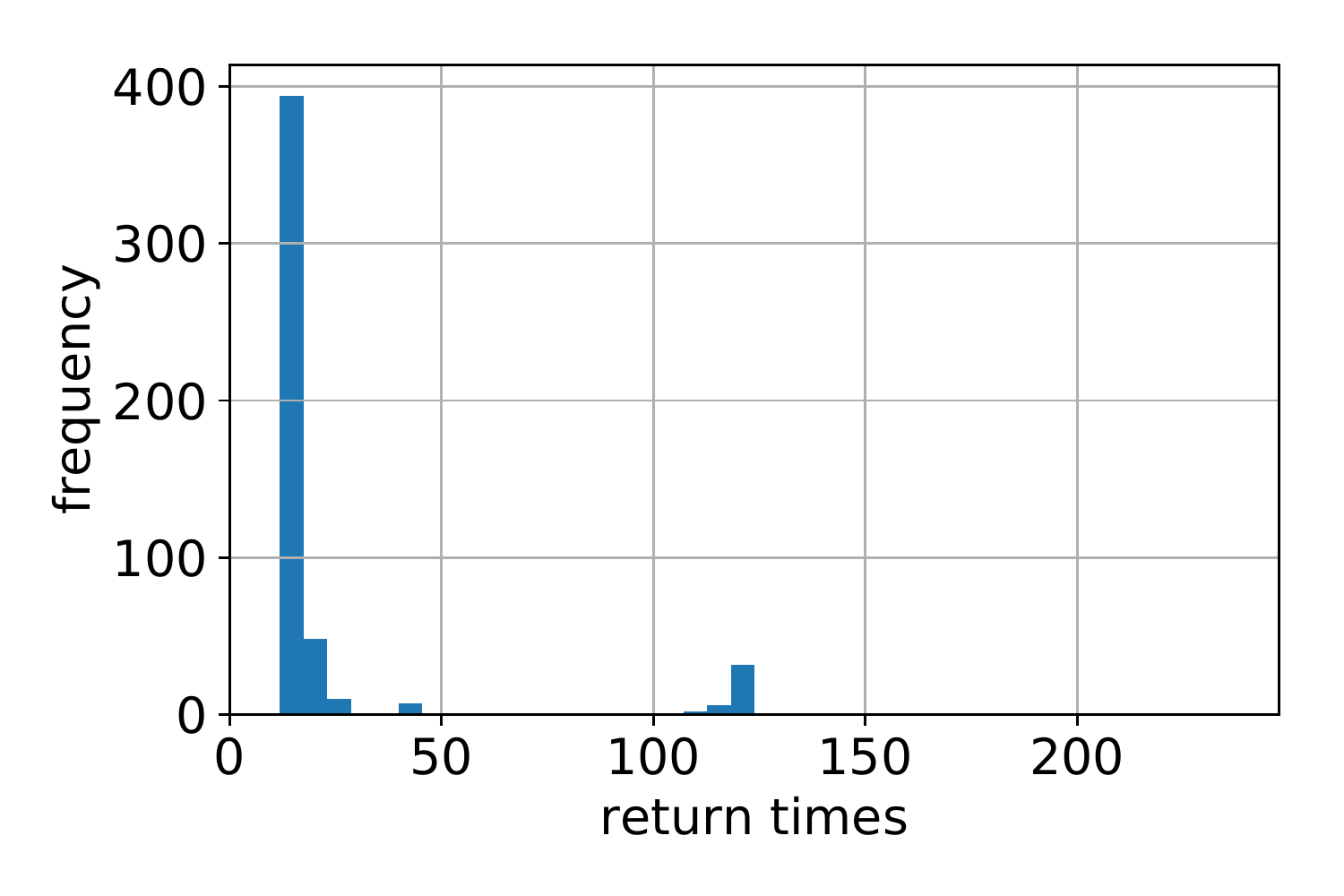}}
  \centerline{(a) AWS Lambda return times}\medskip
\end{minipage}
\hfill
\begin{minipage}[b]{0.47\linewidth}
  \centering
  \centerline{\includegraphics[width=4.5cm]{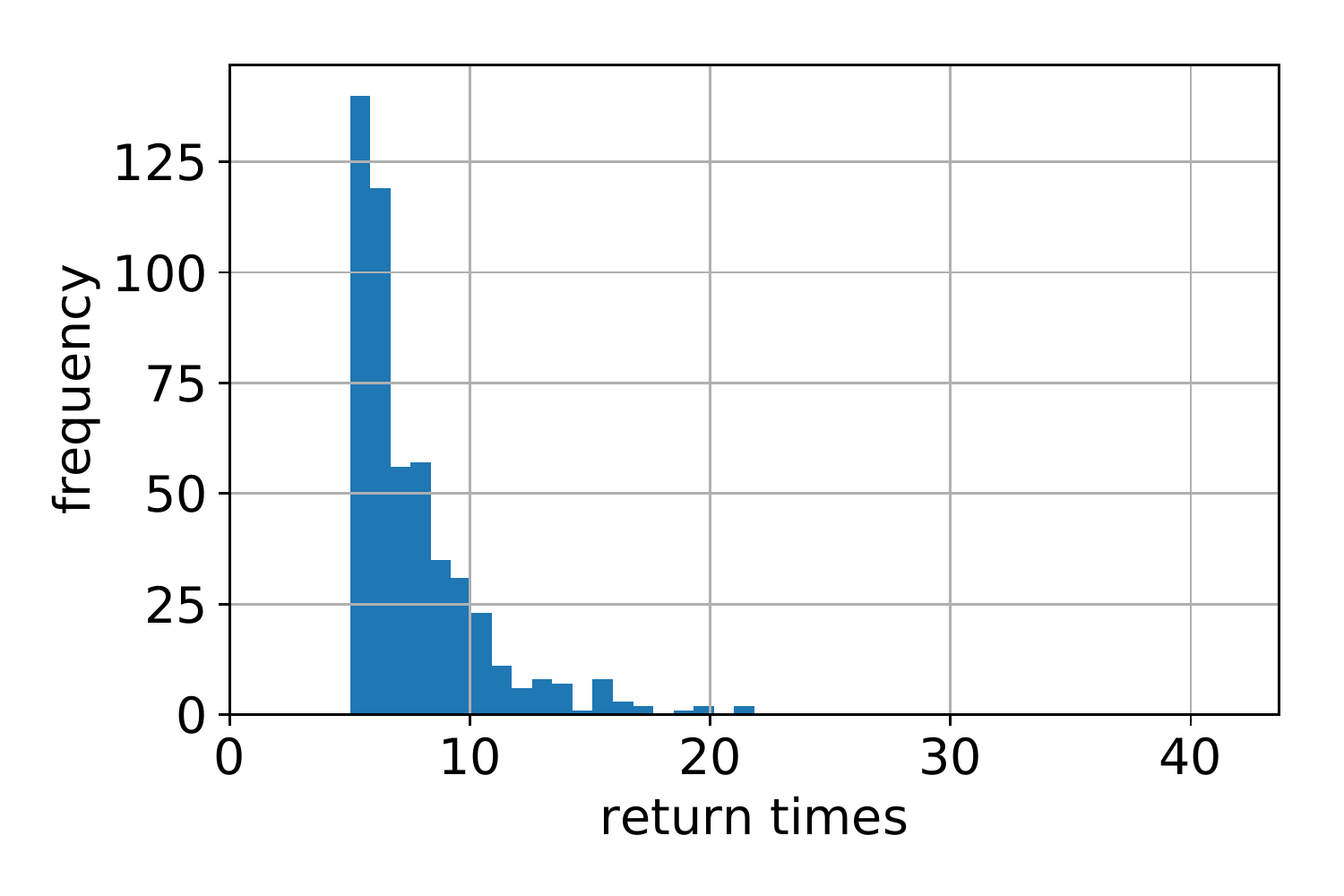}}
  \centerline{(b) Exponential distribution}\medskip
\end{minipage}
\caption{Run time distributions used in simulations.}
\label{run_time_distributions}
\end{figure}

\begin{figure}[htb]
\begin{minipage}[b]{0.47\linewidth}
  \centering
  \centerline{\includegraphics[width=4.4cm]{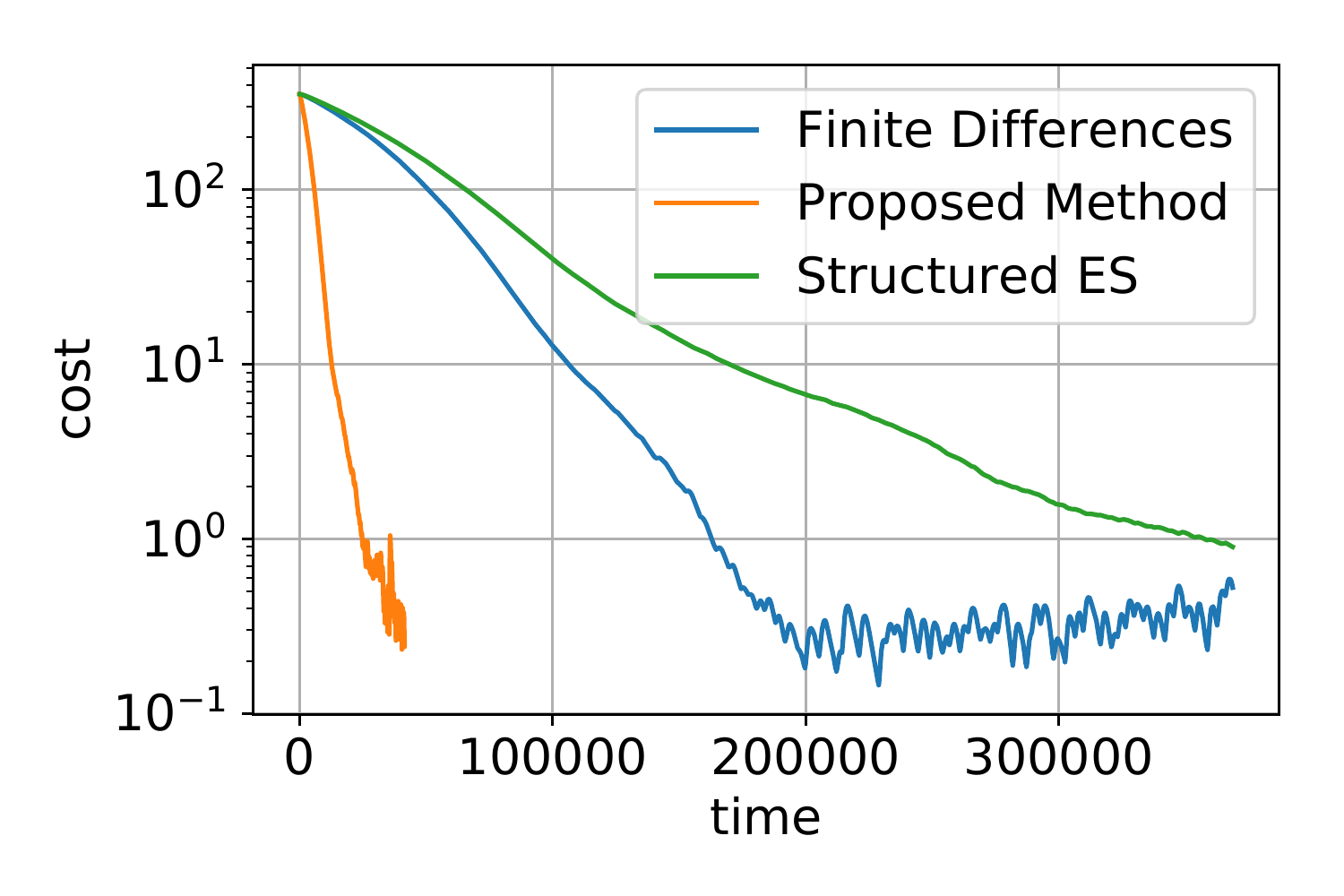}}
  \centerline{(a)}\medskip
\end{minipage}
\hfill
\begin{minipage}[b]{0.47\linewidth}
  \centering
  \centerline{\includegraphics[width=4.4cm]{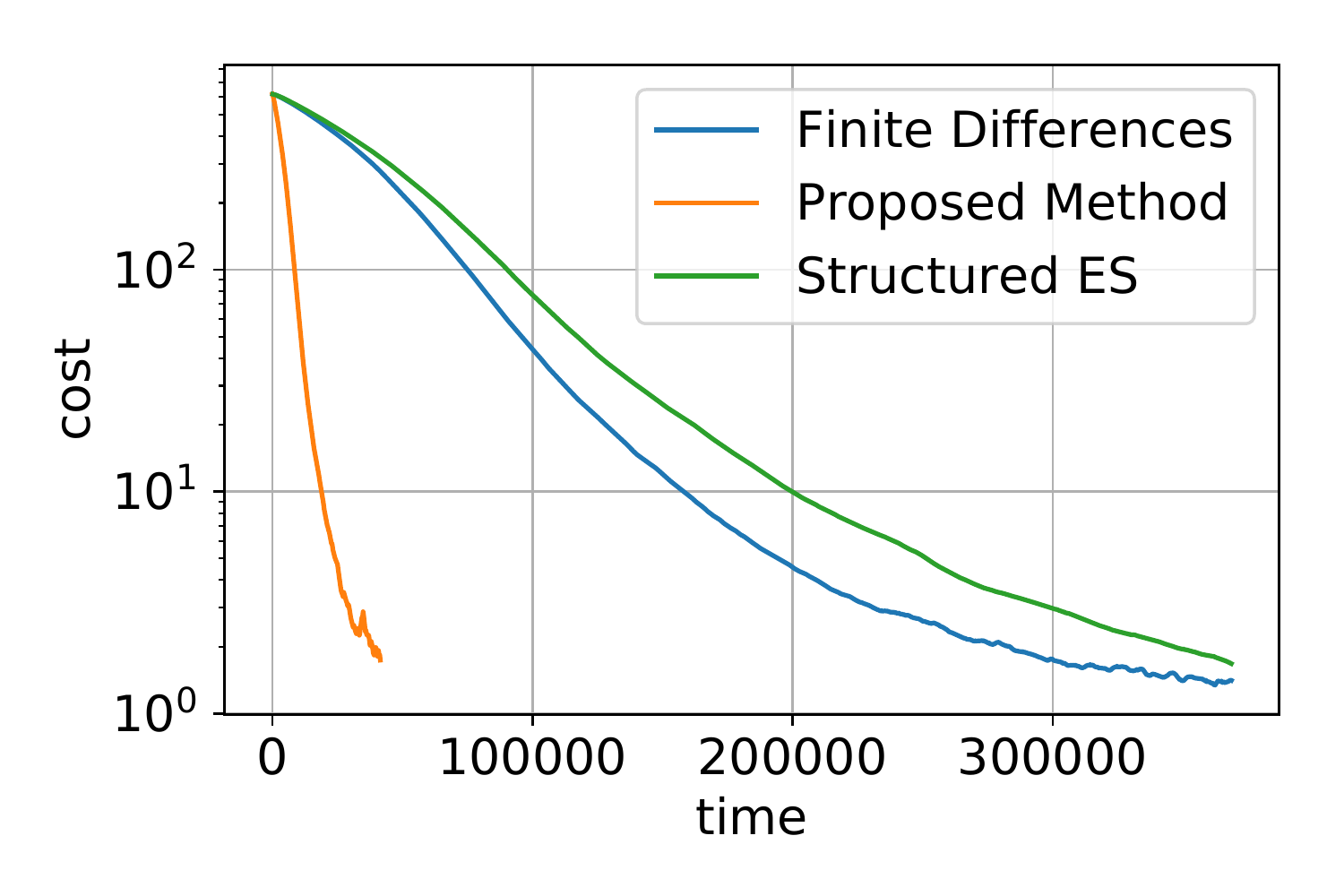}}
  \centerline{(b)}\medskip
\end{minipage}
\hfill
\begin{minipage}[b]{.47\linewidth}
  \centering
  \centerline{\includegraphics[width=4.4cm]{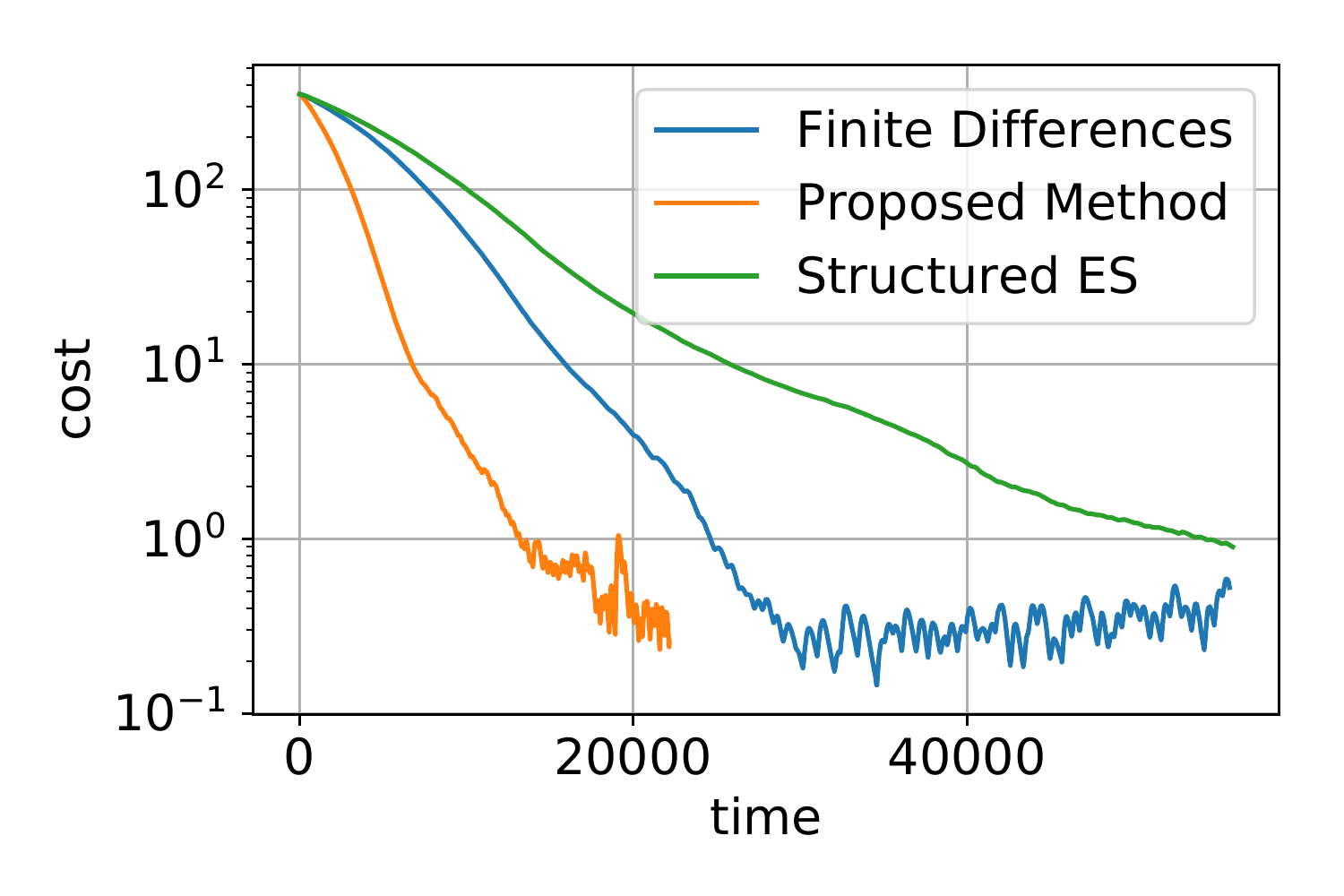}}
  \centerline{(c)}\medskip
\end{minipage}
\hfill
\begin{minipage}[b]{0.47\linewidth}
  \centering
  \centerline{\includegraphics[width=4.4cm]{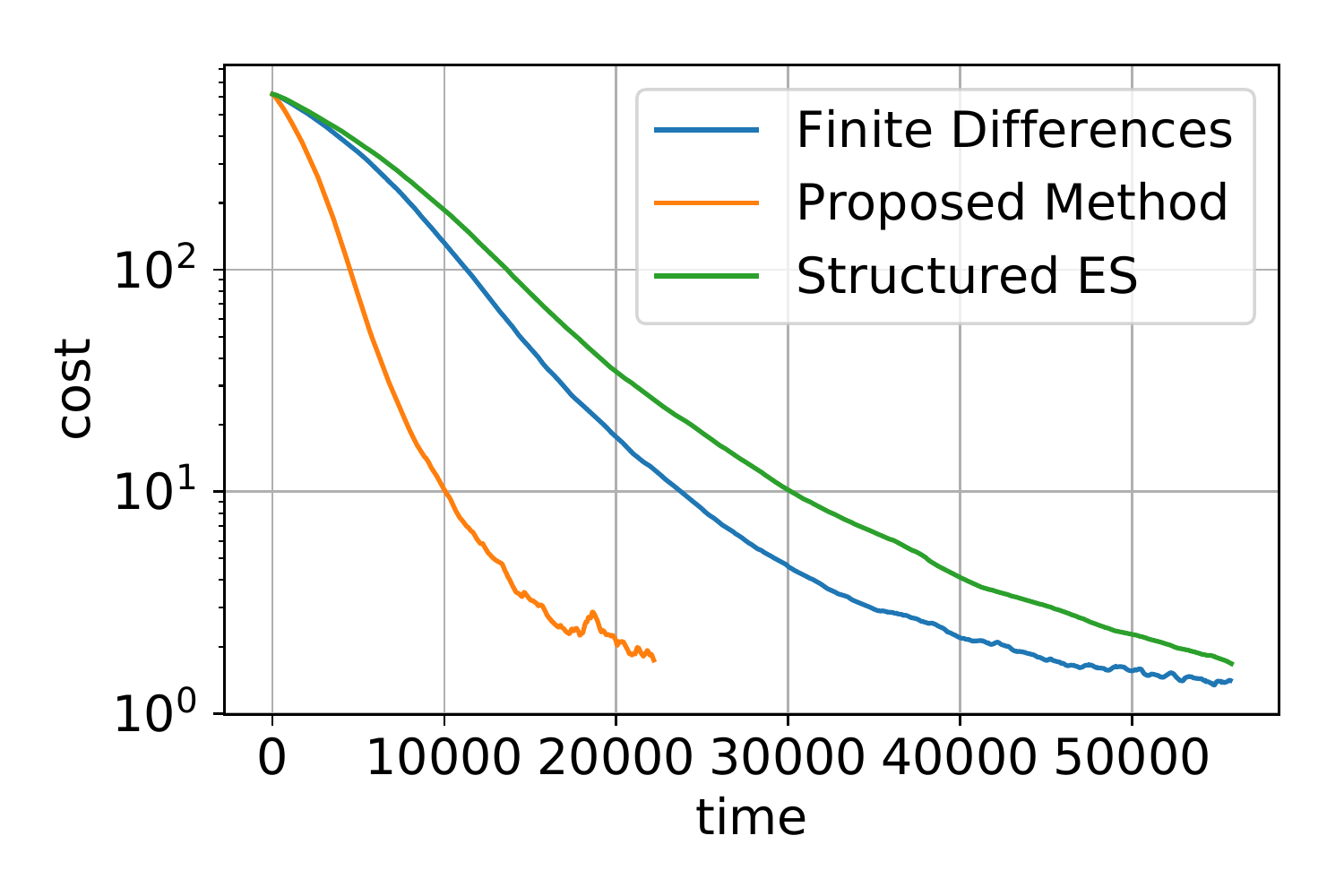}}
  \centerline{(d)}\medskip
\end{minipage}
\caption{Simulated plots of cost as a function of time (all methods were run for 3000 iterations). a,c: Airplane image classified as dog. b,d: Ship image classified as dog. a,b: Worker return times sampled from AWS Lambda return times shown in Figure \ref{run_time_distributions}(a). c,d: Worker return times sampled from the exponential distribution shown in Figure \ref{run_time_distributions}(b).}
\label{adversarial_plots}
\end{figure}

We now discuss details of how we obtained these adversarial images. For obtaining these results, we have solved the problem of minimizing $f(\theta)$ where $f(\theta)$ is given in \eqref{f_targeted}. We have set $\kappa=0$, $c=0.1$. We have solved this optimization problem using three different methods and compared them in Figure \ref{adversarial_plots} via a run time simulation. The simulation was done by sampling worker run times from the distributions shown in Figure \ref{run_time_distributions}. Figure \ref{run_time_distributions}(a) shows the empirical histogram of worker times which were obtained for the AWS Lambda platform, which is a serverless computing platform. Figure \ref{run_time_distributions}(b) shows the histogram of worker times generated from the shifted exponential distribution.

Figure \ref{adversarial_plots} illustrates that it is possible to eliminate the straggling workers effect by using the proposed method. The way finite differences and structured evolution strategies methods were implemented is not straggler-resilient in these simulations, that is, all worker outputs are waited for in every gradient update. We note that this is different from the results on the optimization of the $l_1$-norm based function given in previous subsection because we used straggler-resilient versions of the methods in those experiments. The point of the simulation results given in Figure \ref{adversarial_images} is not necessarily to compare the run times of each method but instead to showcase that our proposed method works for the recent problem of adversarial attacks to deep neural networks.

\section{Conclusion}
We have introduced a novel distributed black-box optimization mechanism based on coded perturbation directions. Because of the redundancy inserted during encoding, the proposed method is resilient against straggling workers in distributed computing platforms. This makes our method suitable for large scale serverless cloud-based systems where nodes are unreliable. Our work utilizes coding theoretic tools and ideas from coded distributed computation methods. Furthermore, the proposed method provides a connection between two black-box optimization methods: finite differences and structured evolution strategies, in a way that it is possible to go from one to the other and choose the one that drives the cost down more effectively. We have provided an argument why coded gradient estimates would work for a variety of cost functions and shown this through numerical examples.

\bibliographystyle{plain}
\bibliography{mert}

\end{document}